\documentclass[12pt]{article} 
\usepackage{amsmath}
\usepackage{amsfonts}
\usepackage{times}
\usepackage{epsf}
\def\({\left(}
\def\){\right)}
\def\[{\left[}
\def\]{\right]}
\def\non{ \nonumber }

\def\ga{\gamma}

\def\Q{\mathcal{Q}}
\def\z{\zeta}
\def\ga{\gamma}
\def\debut{\begin{align}}
\def\fin{\end{align}}
\def\half{\textstyle{\frac 1 2}}
\def\Res{\mathcal{R}}
\begin{document}
\rightline{LPTHE--03-06}
\
\vskip 1cm
\centerline{\LARGE Baxter equations and Deformation of }
\bigskip
\centerline{\LARGE  Abelian Differentials.}
\vskip 2cm
\centerline{\large 
F.A. Smirnov 
\footnote[0]{Membre du CNRS}}
\vskip1cm
\centerline{  Laboratoire de Physique Th\'eorique et Hautes
Energies \footnote[1]{\it Laboratoire associ\'e au CNRS.}}
\centerline{ Universit\'e Pierre et Marie Curie, Tour 16 1$^{er}$
		\'etage, 4 place Jussieu}
\centerline{75252 Paris cedex 05-France}
\vskip2cm
\noindent
{\bf Abstract.} 
In this paper the proofs are given of important properties of 
deformed Abelian differentials introduced earlier in connection
with quantum integrable systems. The starting point of the
construction is Baxter equation. In particular, we prove 
Riemann bilinear relation. Duality plays important role in
our consideration. Classical limit is considered in details.

\newpage
\section{Introduction.}

It is well known that spectra of quantum integrable models are defined
by solutions of Baxter equations \cite{bax}. To our mind  most naturally
these equations appear in the method of separation of
variables developed by Sklyanin \cite{skl}. In the classical case 
the separation of variables is closely related with
algebra-geometrical methods based on spectral curves and their
Jacobi varieties.One can think of Baxter equations as of definition
of ``quantum spectral curves''.

The averages of observables in the quantum
case are written in terms of rather peculiar integrals.
In this paper we shall consider the case related to $U_q(sl _2)$ in which
all these integrals can be expressed in terms of one-fold deformed
hyper-elliptic integrals. In classical case this statement corresponds
to rather non-trivial property of cohomologies of
affine hyper-elliptic Jacobian which was conjectured in \cite{sn}
and proved in \cite{n}.
These deformed hyper-elliptic integrals in are considered in the paper \cite{qaf},
but the details of proof of their properties have
never been given. This will be done in the present paper because we
believe them to be interesting and instructive.

Let us be more specific  in discussion of Baxter equations.
If one considers a quantum integrable system with finitely many
degrees of freedom related to $U_q(sl _2)$ the generating function of
its integrals of motion is given by a polynomial: 
\begin{align}
&t(z)=\sum\limits _{k=0}^{g+1}z^kt_k\label{t}
\end{align}
where $z$ is the spectral parameter and the
number of degrees of freedom equals $g$ and $t_0=1$. 
If we consider a system
of $U_q(sl _2(\mathbb{R}))$ type the spectrum is defined
by solutions to Baxter equations:
\begin{align}
&\Q(\z +i\ga )+\Q(\z -i\ga )=t(z)\Q(\z )
\label{bax1}
\end{align}
where $\gamma$ is the coupling constant (or Plank constant), 
$\zeta =\frac 1 2 \log (z)$, $\Q(\z )$ is an entire function
with certain requirements on position of its zeros and the
following asymptotic:
\begin{align}
& \Q (\z )=e^{-\frac{g+1}{2\ga}((\pi+\ga)\z\pm 2i\z^2)}
\label{ass}
\end{align}
if $\z\to\infty$ being not too far in upper (lower) half-planes. 

There is a nice argument due to Al. Zamolodchikov 
\cite{alz} which explains
the appearance of duality in this situation. Notice that $\Q(\z +\pi i ) $
is also solution to the same equation (\ref{bax1}). The quantum Wronskian of
these two solutions must be entire $i\gamma$-periodical function,
but due to the asymptotic (\ref{ass}) it equals one:
\begin{align}
\Q\(\z +\frac{\pi +\ga}2\ i \)\Q\(\z -\frac{\pi +\ga}2\ i \)-
\Q\(\z +\frac{\pi -\ga}2\ i \)\Q\(\z -\frac{\pi -\ga}2\ i \)=1
\label{Wr}
\end{align}
It is easy to see that (\ref{Wr}) together with the asymptotic
(\ref{ass}) actually imply the existence of
polynomial $t(z)$ with which the Baxter equation (\ref{bax1}) holds.
Indeed, take a function $\Q(\z )$ satisfying 
(\ref{Wr}) and define
$$ t(z)=\frac{\Q(\z +i \ga) +\Q(\z -i\ga) }
{\Q(\z )}$$
from (\ref{Wr}) it is easy to see that this is indeed a function
of $z$ ($\pi i$-periodical function of $\z$) without singularities, and (\ref{ass})
allows to show that $t(z)$ is in fact a polynomial of degree $g+1$.

But two periods: $i\gamma$ and $i\pi$ enter (\ref{Wr}) in completely
symmetric way. So, the same kind of reasonings proves existence of another
polynomial of degree $g+1$ which we denote by $T(Z)$ and with which the
dual Baxter equation is satisfied:
\begin{align}
&\Q(\z +i\pi )+\Q(\z -i\pi )=T(Z)\Q(\z )
\label{bax2}
\end{align}
here $ Z=\exp\(\frac {2\pi}{\gamma}\z\)$.
The pair of dual equations (\ref{bax1}) and (\ref{bax1}) 
and mathematical structures related to the is subject of this paper.

The duality considered in this paper has much in common with the
duality for representations of $U_q(sl_2(\mathbb{R}))$ \cite{pt}.
Actually, the type of real form for integrable models that we have
in mind is similar in the classical case to taking $SL_2(\mathbb{R})$ 
as real form of $SL_2(\mathbb{C})$.
\newline
{\bf Acknowledgments.} This work is partly supported by 
INTAS grant 00-00055. The author is grateful to organizing
committee 
of  "6th International Workshop
Conformal Field Theory and Integrable Models " in Chernogolovka
for warm hospitality.

\section{Deformed Abelian differentials.}

In this section we give formal definition concerning
deformed Abelian differentials. The properties of these
differentials are discussed in next sections.

Consider a solution $\Q(\z )$  to Baxter equation:
\begin{align}
&\Q(\z +i\ga )+\Q(\z -i\ga )=t(z)\Q(\z )
\label{bax}
\end{align}
Asymptotically it behaves as follows:
\begin{align}
&\Q (\z)=\Q_+(\z )+\Q_-(\z ),\non\\
& \Q _{\pm}(\z )=e^{\pm (g+1)\frac{\z^2}{i\ga}}
(zZ)^{-\frac {g+1} 2}
f_{\pm }(z)F_{\pm}(Z)\non
\end{align}
In this section we shall consider:
$$ 
\varphi _{\pm}(\z )=
e^{\pm (g+1)\frac{\z^2}{i\ga}}
z^{-\frac {g+1} 2}
f_{\pm }(z)
$$
which are 
formal
solutions to  the equations
$$\varphi _{\pm}(\z+i\ga  )+
\varphi _{\pm}(\z -i\ga)
=(-1)^{g+1}t(z)\varphi _{\pm}(\z )$$
These asymptotic solutions satisfy the q-Wronskian
relation:
\begin{align}
&\varphi _+(\z +i\ga)\varphi _-(\z )-
\varphi _-(\z +i\ga)\varphi _+(\z )=1
\end{align}

In this section we shall use the space $\text{w}$ of
semi-infinite Laurent series of the form
$$a(z)=\sum\limits _{k=-\infty}^N a_kz^k$$
where $N$ is arbitrary but finite.
Let us give some definitions.
\newline
{\bf Definition 1.} For Laurent series $a(z)$
define:
\begin{align}
\Res (u)=\text{res} \( z^{-1}a(z)\varphi _+(\z)\varphi _-(z)\)
\label{Res}
\end{align}

The expression 
in RHS is Laurent series in $z$, so, the 
residue is well defined.
Our next goal is to define a pairing between Laurent
series without residues.
\newline
{\bf Definition 2.}
Consider $a,b\in \text{w}$ such that
$\Res (a)=\Res (a)=0$ then
\begin{align}
a\circ b=i\ga \ \text{res}\bigl( a(z)z^{-1}\bigl(
2&\varphi _+(\z)\varphi _-(\z)\delta _{\ga}^{-1}
(b(z)\varphi _+(\z)\varphi _-(\z))-\non\\
-
&\varphi _+(\z)\varphi _+(\z)\delta _{\ga}^{-1}
(b(z)\varphi _-(\z)\varphi _-(\z ))-\non\\-&
\varphi _-(\z)\varphi _-(\z)\delta _{\ga}^{-1}
(b(z)\varphi _+(\z)\varphi _+(\z))\bigr)
\bigr)
\label{pair}
\end{align}
Where we introduced the following notation:
$\forall \eta\in \mathbb{R}$ define
$$\delta _{\eta}(f(\z))=f(\z+i\eta )-f(\z )$$
Definition 2 requires some comments. 

First, since
$b(z)\varphi _+(\z)\varphi _-(\z)$ is Laurent series and
$\Res (b)=0$ the expression 
$\delta _{\ga}^{-1}
(b(z)\varphi _+(\z)\varphi _-(\z))$ is well-defined as Laurent
series up to a constant. The latter ambiguity does not affect
the definition since $\Res (a)=0$. 

Second, the meaning of  
$\delta _{\ga}^{-1}
(b(z)\varphi _-(\z)\varphi _-(\z))$ and
$\delta _{\ga}^{-1}
(b(z)\varphi _+(\z)\varphi _+(\z))$
must be clarified. Notice that 
$$b(z)\varphi _{\pm}(\z)\varphi _{\pm}(\z)
=e ^{\pm 2(g+1)\frac {\z ^2}{i\ga }}v_{\pm}(z)$$
where $v_{\pm} (z)$ are Laurent series. We define
$$\delta _{\ga}^{-1}
(b(z)\varphi _{\pm}(\z)\varphi _{\pm}(\z))=
e ^{\pm 2(g+1)\frac {\z ^2}{i\ga }}u_{\pm}(z)$$
where
$$(qz)^{2(g+1)}u_+(zq^2)-u_+(z)=v_+(z),\quad
(qz)^{-2(g+1)}u_-(zq^{2})-u_-(z)=v_-(z)
$$
Obviously the $u_{\pm}(z)$ are well-defined as Laurent series.

It is easy to see that the pairing (\ref{pair}) is skew-symmetric:
$$a\circ b=-b\circ a$$

Our next definition introduces "exact forms".
\newline
{\bf Definition 3.}
For any $a(z)\in\text{w}$  define
\begin{align}
(\mathcal{D}a)(z)&=\frac 1 {i\ga }\(t(z)\pi _q (t(z)a(z))
+a(zq^{-2})-a(zq^2)\)
\end{align}
where 
$$ \pi _q\(\sum b_kz^k\)=\sum \frac {q^{2k}-1}
{q^{2k}+1}\ b_k z^k
$$

The importance of the definition of $\mathcal{D}$ is due to
the following fact:
\begin{align}
(\mathcal{D}a)(z)\varphi _{\epsilon }(\z)\varphi _{\epsilon '}(\z)=
\delta  _{\gamma }(x_{\epsilon ,\epsilon '}(z))\label{exact}
\end{align}
for  any $\epsilon$, $\epsilon '$ equal $+$ or $-$, the function
 $x_{\epsilon ,\epsilon '}(z)$ is given by
\begin{align}
x_{\epsilon ,\epsilon '}(z)
=\frac 1{2i\ga}&\(
(-1)^g \(\pi _q(a(z)t(z))-a(z)t(z)\)\right.
\non\\ &\times
(\varphi _{\epsilon }(\z-i\ga)\varphi _{\epsilon ' }(\z)
+\varphi _{\epsilon }(\z)\varphi _{\epsilon '}(\z-i\ga))-\non\\
&\left.
-2a(zq^{-2})\varphi _{\epsilon }(\z)\varphi _{\epsilon '}(\z)
-2a(z)\varphi _{\epsilon }(\z-i\ga)\varphi _{\epsilon '}(\z-i\ga)
\right)
\non
\end{align}
Since $x_ {+,-}(z)$ are Laurent series we have, in particular,
$$\Res (\mathcal{D}(a))=0$$

Now we introduce important object: the space of deformed Abelian
differentials.
Consider the space $\text{v}$ of polynomials of one
variable $a(z)$ such that $a(0)=0$. 
These polynomials will define deformed Abelian differentials
on the affine hyper-elliptic curve.

Consider a polynomial $u(z)$ which does not necessarily
vanishes at $z=0$. It is easy to see
that $b=\mathcal{D}u\in \text{v}$. 
Let us calculate  $a\circ b$
for $a\in \text{v}$. Simple calculation gives:
\begin{align}
a\circ b= &\text{res}\(z^{-1} a(z)u(z)
\(\varphi _+(\z)\varphi _-(\z-i\ga)-
\varphi _-(\z)\varphi _+(\z-i\ga)\)^2\)=\non\\
=&\text{res}\(z^{-1} a(z)u(z)\)
=0\label{expair}
\end{align}
So the pairing between $\text{v}$ and exact form vanishes.


Let us introduce the following basis in the space $\text{v}$:
\begin{align}
&r_k(z)=z^{k},\quad  \text{for}\ k=1,\cdots g+1,\non\\
\ \non\\
&s_k(z)=\frac 1 {i\ga}\(
t(z)\pi _q\bigl(\[ t(z)z^{-k}\]_>\bigr)+[z^{-k}]_>(q^{2k}-q^{-2k})\),
\quad\text{for} \ -\infty < k\le g+1\non
\end{align}
where $[\cdots ]_>$ means the positive degree part of
Laurent series in brackets. 
Obviously,
$$\text{deg}(r_k)=k,\quad\text{deg}(s_k)=2g+2-k$$
so, all the degrees are taken into account. Notice also
that there is only one polynomial of degree $g+1$
(this is $r_{g+1}(z)$) because $s_{g+1}(z)=0$.
We shall call the polynomials $r_k$ for $k=1,\cdots , g$
first kind, $s_k$ for $k=1,\cdots , g$ second kind,
$r_{g+1}$ - third kind. The polynomials $s_k$ for $k\le 0$
are exact forms. 

Introduce the Laurent polynomial 
\begin{align}
s_k^-(z)=\frac 1 {i\ga}\(t(z)\pi _q\bigl(\[ t(z)z^{-k}\]_<\bigr)+
[z^{-k}]_<(q^{2k}-q^{-2k})\)
\label{s-}
\end{align}
Where $[\cdots ]_<$ means that only negative part
of Laurent series is taken.
Evidently,
\begin{align}
s_k(z)+s^-_k(z)=\mathcal{D}(r_{-k})(z),
\quad r_{-k}(z)=z^{-k}
\label{s+s-}
\end{align}

Consider the residues of our polynomials. It is clear that
for $z\to\infty$
$$r_k(z)\ \varphi _+(\z)\varphi _-(\z)=o(1),\qquad 
\text{for}\  k=1,\cdots ,g$$
so, these $r_k$ have no residues. It is equally clear that
$$\Res (r_{g+1})=1$$
as it should be for the third king differential. Finally,
notice that
$$s^-_k(z)\ \varphi _+(\z)\varphi _-(\z)=o(1)$$
so, from (\ref{s+s-}) it follows that 
$$\Res (s_k)=0$$

Consider the pairing between those polynomials whose
residues vanish.
It is easy to show that 
$$r_k\circ r_l=0$$
for $k,l =1,\cdots, g$ just because $r_k$ do not grow
sufficiently fast.
Now let us calculate $a\circ s_k$ for arbitrary $a$.
Using the formula (\ref{s+s-}) one finds 
\begin{align}
a\circ s_k= \text{res}\(a(z)z^{-k-1}\)-a\circ s_k^-\label{ask}
\end{align}
If $a=r_l$ the second term in (\ref{ask}) vanishes because
$a(z)$ and $s_k(z)$ do not grow sufficiently fast. So, one finds
$$ r_l\circ s_k =\delta _{kl}$$
Consider the case 
$a=s_l$. Using
$$s_l\circ s^-_k=-s^-_k\circ s_l$$
find
\begin{align}
s_l\circ s_k=\text{res}\(s_l(z)z^{-k-1}
+
s_k^-(z)z^{-l-1}\)+ s_l^-\circ s_k^-\label{slsk}
\end{align}
The last term in the RHS vanishes because $s_l^-, s_k^-$
do not grow fast enough.
It is a nice exercise to show that 
the first term in the RHS of
(\ref{slsk}) vanishes using explicit formulae
for $s_l$ and $s_k^-$. 

\section{Deformed Abelian integrals.}
In this section we shall define deformed Abelian
integrals. Let us start from the simplest case.
Consider the polynomials
$$R_l(Z)=Z^{l}$$ 
for $l=1,\cdots ,g$. Take a polynomial
$a(z)\in  \text{v}$. Suppose that $\gamma $ is sufficiently
small, more precisely, we assume that
$$\text{deg}(a(z))<\frac {\pi}{\gamma}$$
Then the following
integral is well defined:
\begin{align}
&\langle a,R_l\rangle\equiv\int\limits _{-\infty}^{\infty}a(z)R_l(Z)
\ \Q (\z )^2d\z
\label{<>}
\end{align}
Our first goal is to define $\langle a,R_l\rangle$ for arbitrary $\gamma$.
We shall use the basis $r_k$ ($k=0,\cdots , g$),
$s_k$ ($k\le g+1$) introduced in the previous section.

If we substitute $a=r_k$  into (\ref{<>})
the integral converges for any $\ga$, but for $a=s_k$
a regularization is needed. 

In what follows we shall use the operators $w$, $w^*$ which
act as usual:
\begin{align}
&w\Q(\z)=\Q(\z-i\ga),\qquad
\Q(\z)w^*=\Q(\z-i\ga),\non
\end{align}

The following important identity holds:
\begin{align}
&s_k(z)\ \Q(\z)^2=
\Q(\z)\widetilde{s}_k(z,w)\Q(\z)+
\delta _{\gamma}\(\frac 1 {i\ga}
(id
-\pi _q)([t(z)z^{-k}]_>)
\Q(\z)w\Q(\z)\)
\non
\end{align}
where we have introduced new object:
\begin{align}
&\widetilde{s}_k(z,w)=\frac 1 {i\ga}\([z^{-k}]_>
(q^{2k}-q^{-2k})
+
\[z^{-k}t(z)\]_>\(2w-t(z)\)\)\non
\end{align}
Notice that if $\ga $ is sufficiently small  the integral (\ref{<>})
can be rewritten in the following way
$$\langle s_k,R_l\rangle=\int\limits _{-\infty}^{\infty}\Q(\z)R_l(Z)
\widetilde{s}_k(z,w)\Q(\z)d\z$$
because $s_k\Q^2$ 
and $\Q\ \widetilde{s}_k\Q$ differ by
"total difference" and $R_l(Z)$ is $i\ga$-periodical, the "boundary
term" does not contribute for small enough $\ga$.  Let us use this
form of integral in order to define the regularization for arbitrary $\ga$.

For $l=1,\cdots ,g$ consider the integral 
\begin{align}
&\langle s_k,R_l\rangle =\int\limits _{-\infty} ^{\Lambda}\Q(\z)R_l(Z)
\widetilde{s}_k(z,w)\Q(\z)d\z+\label{sR}\\&
+\int\limits _{\Lambda} ^{\Lambda +i\ga}\Q(\z)R_l(Z)
p_k(z,w^*,w)\Q(\z)d\z
-\int\limits _ {\Lambda}^{\infty}\Q(\z)R_l(Z)
\widetilde{s}_k^{\ -}(z,w)\Q(\z)d\z\non
\end{align}
where 
\begin{align}
&\widetilde{s}_k^{\ -}(z,w)=\frac 1 {i\ga}\(
[z^{-k}]_<(q^{2k}-q^{-2k})+
\[z^{-k}t(z)\]_{<}\(2w-t(z)\)\),
\non\\
&p_k(z,w^*,w)=\frac 1 {i\ga}\(z^{-k}\(w^*w+q^{2k}\)
-t_kw\)\non
\end{align}
The properties of the integrals in the RHS of (\ref{sR}) are:
\newline
1. The third integral in (\ref{sR}) converges $\forall \ga$.
Indeed, it is easy to show that the integrant decreases
exponentially as $\z\to\infty$ for all $1\le l\le g+1$. 

\noindent
2. The RHS of (\ref{sR}) does not depend on $\Lambda $ due to the identity:
\begin{align}
\delta _{\gamma}\bigl(&\Q(\z)p_k(z,w^*,w)\Q(\z)\bigr)=
-\Q(\z)\( \widetilde{s}_k(z,w)+\widetilde{s}_k^{\ -}(z,w)  \) \Q(\z)
\non
\end{align}
3. When $\gamma$ is sufficiently small 
and  $1\le l\le g$ one can take the limit
$\Lambda\to\infty$ reproducing the original definition.
In the case $l=g+1$ the regularization
is needed for any $\ga$, so, our definition is not founded
independently, however, an important evidence of self-consistency
will follow from Riemann bilinear relation.
\newline
4. In the original definition $s_{g+1}(z)=0$. Consider, however,
the definition of regularized integral. It does not
vanish because of contributions 
$\int\limits _{\Lambda}^{\Lambda +i\ga}$ and
$\int\limits _{\Lambda}^{\infty}$. The integrals can be easily 
evaluated:
\begin{align}
&\langle s_{g+1}, R_l\rangle = \ \delta _{l,g+1}\non
\end{align}

Thus the formula (\ref{sR}) provides  an analytical continuation of 
the original definition (\ref{<>}) for arbitrary $\ga$. Certainly to affirm that
we have to assume that the solution to Baxter equations $\Q(\z)$
allow analytical continuation with respect to $\ga$. 

Similarly to $s_k$ consider
\begin{align}
S_k(Z)=\frac 1{i\pi}\(T(z)\pi _Q\([T(z)Z^{-k}]_> \)
+\[Z^{-k}\]_>(Q^{2k}-Q^{-2k})\)\non ,
\end{align}
For these dual objects define:
\begin{align}
&\langle r_l,S_k\rangle =\int\limits _{-\infty} ^{\Lambda}\Q(\z)r_l(z)
\widetilde{S}_k(Z,W)\Q(\z)d\z+\label{rS}\\&
+\int\limits _{\Lambda} ^{\Lambda +i\pi}\Q(\z)r_l(z)
P_k(Z,W^*\hskip -.1cm,W)\Q(\z)d\z
-\int\limits _ {\Lambda}^{\infty}\Q(\z)r_l(z)
\widetilde{S}_k^{\ -}    ( Z,W)        \Q(\z)d\z\non
\end{align}
where
\begin{align}
&\widetilde{S}_k  ( Z,W)\ \   =\ \ \ \frac 1{i\pi}\(
\ \[Z^{-k}\]_>(Q^{2k}-Q^{-2k})+
\[Z^{-k}T(Z)\]_{>}\(2W-T(Z)\)\),\non\\
&\widetilde{S}_k^{\ -}    ( Z,W)\ \  =\ \ \frac 1{i\pi}\(
\[Z^{-k}\]_{<}(Q^{2k}-Q^{-2k})+
\[Z^{-k}T(Z)\]_{<}\(2W-T(Z)\)\),\non\\
&P_k(Z,W^*\hskip -.1cm,W)=\frac 1{i\pi}\(Z^{-k}\(W^*W+Q^{2k}\)
-T_kW\)\non
\end{align}

Now we want to define the pairing $\langle s_k,S_l\rangle$.
Notice that even in the region of small $\ga $ corresponding
integral is not defined in a simple way. So, we shall define this
pairing by analogy, the real justification of our definition
will be provided later by  Riemann bilinear relation.

Define
\begin{align}
&\langle s_k,S_l\rangle=
\int\limits _{-\infty} ^{\Lambda}\Q(\z)
\widetilde{s}_k(z,w)
\widetilde{S}_l(Z,W)
\Q(\z)d\z+\label{sS1}\\
&+\hskip -.2cm\int\limits _{\Lambda } ^{\Lambda  +i\ga}
\hskip -.2cm\Q(\z)
\widetilde{S}_l(Z,W)p_k(z,w^*\hskip -.1 cm ,w)\Q(\z)d\z-
\int\limits _{\Lambda}^{\Lambda '}\Q(\z)\widetilde{s}_k^{\ -}(z,w)
\widetilde{S}_l(Z,W)
\Q(\z)d\z-\non\\
&-\hskip -.2cm
\int\limits _{\Lambda'}^{\Lambda '+i\pi}\hskip -.2cm\Q(\z )
\widetilde{s}_k^{\ -}(z,w)
P_l (Z,W^*\hskip -.1cm ,W)\Q(\z)d\z 
+\int\limits _ {\Lambda '}^{\infty}\Q (\z)
\widetilde{s}_k^{\ -}(z,w)
\widetilde{S}_l ^{\ -}(Z,W)\Q(\z)d\z\non
\end{align}
By construction this definition does not really depend
on $\Lambda $, $\Lambda '$.
Duality requires that (\ref{sS1}) is equivalent to the
following one
\begin{align}
&\langle s_k,S_l\rangle=
\int\limits _{-\infty} ^{\Lambda}\Q(\z)
\widetilde{s}_k(z,w)
\widetilde{S}_l(Z,W)
\Q(\z)d\z+\label{sS2}\\
&+\hskip -.2cm
\int\limits _{\Lambda } ^{\Lambda +i\pi}\hskip -.2cm\Q(\z)
\widetilde{s}_k(z,w)P_l (Z,W^*\hskip -.1cm,W)\Q(\z)d\z 
-\hskip -.1cm\int\limits _{\Lambda } ^{\Lambda '}\hskip -.1cm\Q(\z)
\widetilde{s}_k(z,w)
\widetilde{S}_l^{\ -}(Z,W)\Q(\z)d\z -\non\\
&-\hskip -.2cm\int\limits _{\Lambda '} ^{\Lambda ' +i\ga}
\hskip -.2cm\Q(\z)
\widetilde{S}_l^{\ -}(Z,W)p_k(z,w^*\hskip -.1cm,w)\Q(\z)d\z
+\hskip -.1cm\int\limits _ {\Lambda '}^{\infty}\Q(\z)
\widetilde{s}_k^{\ -}(z,w)
\widetilde{S}_l ^{\ -}(Z,W)\Q(\z)d\z\non
\end{align}
Let us show that the equivalence indeed holds.
To this end consider 
\newline
$\Lambda '\to\Lambda $. Then the
equivalence in question requires:
\begin{align}
&\int\limits _{\Lambda } ^{\Lambda  +i\ga}\Q(\z)
\ \widetilde{S}_l(Z,W)p_k(z,w^*\hskip -.1cm,w)\Q(\z)d\z-\non\\
-&\int\limits _{\Lambda } ^{\Lambda +i\pi}\Q(\z)
\ \widetilde{s}_k ^{\ -}(z,w)
P_l (Z,W^*\hskip -.1cm ,W)\Q(\z)d\z =\non\\=
&\int\limits _{\Lambda } ^{\Lambda  +i\pi}\Q(\z)
\ \widetilde{s}_k(z,w)P_l (Z,W^*\hskip -.1cm ,W)
\Q(\z)d\z -\non\\
-&\int\limits _{\Lambda } ^{\Lambda  +i\ga}\Q(\z)
\ \widetilde{S}_l^{\ -}(Z,W)p_k(z,w^*\hskip -.1cm,w)\Q(\z)d\z\non
\end{align}
or, equivalently:
\begin{align}
&\int\limits _{\Lambda } ^{\Lambda  +i\ga}\Q(\z)
\bigl(\widetilde{S}_l(Z,W)+\widetilde{S}_l^{\ -}(Z,W)\bigr)
p_k(z,w^*\hskip -.1cm,w)\Q(\z)d\z=
\non\\=
&\int\limits _{\Lambda } ^{\Lambda  +i\pi}\Q(\z)
\bigl(\widetilde{s}_k(z,w)+\widetilde{s}_k ^{\ -}(z,w)\bigr)
P_l (Z,W^*\hskip -.1cm ,W)\Q(\z)d\z 
\label{s+s}
\end{align}
Consider the expression:
$$X_{k,l}(\z)=-\Q(\z)
\ p_k(z,w^*\hskip -.1cm,w)P_l (Z,W^*\hskip -.1cm ,W)
\ \Q(\z)$$
one has:
\begin{align}
&
\Q(\z)
\bigl(\widetilde{S}_l(Z,W)+\widetilde{S}_l^{\ -}(Z,W)\bigr)
p_k(z,w^*\hskip -.1cm,w)\Q(\z)
=
\delta _{\pi}\(X_{k,l}(\z)\),\non\\
&
\Q(\z)
\bigl(\ \widetilde{s}_k(z,w)+\widetilde{s}_k ^{\ -}(z,w)\ \bigr)
P_l (Z,W^*\hskip -.1cm ,W)\Q(\z)=
\delta _{\ga}\(X_{k,l}(\z)\),\non
\end{align}
So, the equation (\ref{s+s}) follows from the identity:
$$
\(\int\limits _{\Lambda +i\pi } ^{\Lambda  +i\ga+i\pi}
-\int\limits _{\Lambda } ^{\Lambda  +i\ga}\ \)X_{k,l}(\z)=
\(\int\limits _{\Lambda +i\ga} ^{\Lambda  +i\ga +i\pi}
-\int\limits _{\Lambda } ^{\Lambda  +i\pi}\ \)X_{k,l}(\z)
$$

Recall that $s_k$ for $k\le 0$ are "exact", so, we must have
$$\langle s_k, A\rangle =0 ,\ \text{for} \ k\le 0$$ for any
$A$ which can be either $R_l$ os $S_l$.
In the integrals (\ref{sR}, \ref{sS1}) this property is transparent: corresponding
$\widetilde{s}_k^{\ -}\equiv 0$,
hence one can move $\Lambda $
to $-\infty$, and to check that the integral 
$\int\limits  _{\Lambda}^{\Lambda +i\ga}$    vanishes.
Indeed, $p_k$ is regular at $z\to 0$ in this case while $R_l$ vanishes. 
Similarly  $S_k$ for $k\le 0$ are "exact".

Notice also that the integrals for $\langle s_{g+1}, S_l\rangle$
vanish. So, $\langle s_{g+1}, R_{g+1}\rangle$ remains the only
non-zero pairing involving $s_{g+1}$. Certainly, similar
fact holds for $S_{g+1}$: the only non-zero paring involving
$S_{g+1}$ is
$$\langle r_{g+1}, S_{g+1}\rangle =1$$

Finally among $r_k$ ($1\le k\le g+1$), $s_k$ ($-\infty <k\le g+1$)
on the on hand and  $R_k$ ($1\le k\le g+1$), $S_k$ 
($-\infty <k\le g+1$)
on the other we have defined all the pairings except
for $\langle r_{g+1}, R_{g+1}\rangle$.  This pairing
has no quasi-classical limit. Some regularization can be
proposed in order to define it, but the definition is not
unique. Anyway, one can avoid using this badly defined pairing
in applications to integrable models.

\section{Riemann bilinear relation for deformed
Abelian integrals.}

Riemann bilinear relation is the most important 
property of deformed Abelian differentials.
\newline{\bf Theorem.}
Consider two polynomials $a, b\in \text{v}$ 
such that $\Res (a)=\Res (b)=0$. Then
\begin{align}
&\sum\limits _{l=1}^g \bigl(
\langle a, S_l\rangle
\langle b, R_l\rangle 
-\langle  a, R_l        \rangle
\langle  b, S_l         \rangle\bigr)=\ a\circ b
\label{gh}
\end{align}
{\it Proof.}
The polynomials $s_k$ for $k\le 0$ are exact forms,
so, LHS of (\ref{gh}) for them vanishes as it must be.
The polynomial $r_{g+1}$
has residue. 
So, we shall chose 
$a$ and $b$ from $r_1,\cdots r_g$
and $s_1,\cdots, s_g$.  This 
means, in particular, $\text{deg}(a)\le 2g+1$,
$\text{deg}(b)\le 2g+1$.
It is easy to see that for such polynomials the formula
(\ref{pair}) can be simplified:
\begin{align}
a\circ b &=i\ga \ \text{res}\bigl(z^{-1}a(z)\bigl(
2\varphi  (\z) _+\varphi _-(\z)
\delta _{\ga}^{-1}(b(z)\varphi _+(\z)\varphi _-(\z))+\non\\
&+b(z)\bigl(\varphi _+(\z)
\varphi _-(\z)\bigr)^2\bigr)\bigr)\label{pair1}
\end{align}
Obviously this expression is anti-symmetric with respect to
$a\leftrightarrow b$.

For simplicity we consider the range of small $\ga $.
Namely, we shall require
$$ \ga <\frac {\pi}{2g+2}$$ 
Considering this range of $\ga$
simplifies a lot the regularization for 
$\langle  a, S_l         \rangle$. 
Recall that
\begin{align}
&\widetilde{S}_l^-(Z,W)=\frac 1{i\pi}\(\[Z^{-l}\]_{<}
(Q^{2l}-Q^{-2l})+
\[Z^{-l}T(Z)\]_{<}\(2W-T(Z)\)\)\non
\end{align}
When $\z\to +\infty$ the function
$\Q(\z)\widetilde{S}_l^-( Z,W)\Q(\z)$ decreases as $Z^{-1}$,
for small $\ga $ this  decreasing is very fast, so, the
regularization of  $\langle  a, S_l         \rangle$ when
$a=s_k$ can be seriously simplified. Consider the formula
(\ref{sS2}). For above reasons we can drop last two terms in 
(\ref{sS2}) and put $\Lambda '=\infty$. Also we can 
replace $\widetilde{s}_k$ by $s_k$ because they differ
by   "exact form" and corresponding boundary term vanishes
for small $\gamma$.
The result is:
\begin{align}
&\langle s_k,S_l\rangle=
\int\limits _{-\infty} ^{\Lambda}\Q(\z)
s_k(z)
\widetilde{S}_l(Z,W)
\Q(\z)d\z+\label{simpl}\\
&+\int\limits _{\Lambda } ^{\Lambda +i\pi}\Q(\z)
s _k(z)P_l (Z,W^*\hskip -.1cm,W)\Q(\z)d\z -
\int\limits _{\Lambda } ^{\infty}\Q(\z)
s_k(z)
\widetilde{S}_l^-( Z,W)\Q(\z)d\z \non
\end{align}
Hence for small $\ga$ the formula for 
$\langle a,S_l\rangle $ is absolutely the same for
$a=s_k$ and for $a=r_k$. The integral 
$\int\limits _{\Lambda}^{\infty}$ rapidly converges,
so, we can consider the following formula  for 
in the case of small $\ga $:
\begin{align}
\langle a,S_l \rangle=\lim _{\Lambda\to\infty}
&\( \ \int\limits _{-\infty} ^{\Lambda} \Q(\z)a(z)
\widetilde{S}_l(Z,W)
\Q(\z)d\z
+\right. \non\\ 
+&\left. \ \int\limits _{\Lambda } ^{\Lambda +i\pi}\Q(\z)
a(z)
P_l (Z,W^*\hskip -.1cm,W)\Q(\z)
d\z \)\label{pred}
\end{align}
Recall that $R_l(Z)=Z^{l}$, and for small $\ga $
we do not need any regularization in 
$\langle b, R_l\rangle$. Hence the first sum
of $a\circ b$ can be rewritten as follows:
\begin{align}
&\lim_{\Lambda \to\infty}
\int\limits _{-\infty}^{\Lambda}
d\z 
\int\limits _{-\infty}^{\infty}d\z '\  \Q(\z )a(z)b(z')
\sum\limits _{l=1}^g
\widetilde{S}_l(Z,W)R_l(Z')
\Q(\z )\ \Q(\z ')^2\non\\&+
\int\limits _{\Lambda}^{\Lambda+i\pi}
d\z 
\int\limits _{-\infty}^{\infty}
d\z '\Q(\z ) a(z)b(z')
\sum\limits _{l=1}^gP_l (Z,W^*\hskip -.1cm,W)R_l(Z')
\Q(\z)\ \Q(\z ')^2\label{inte}
\end{align}
Let us evaluate the the sum in the first integrand 
using
$$\sum\limits _{l=1}^g(Z')^{l}\[Z^{-l}T(Z)\]_>=
\frac{Z'T(Z)-ZT(Z')}{Z-Z'}+T_{0}$$
In writing down the result we shall use the notation:
$$\Q(\epsilon _1\ \epsilon _2\ \epsilon_3\ \epsilon_4)=
\Q(\z +i\epsilon _1\pi)\Q (\z +i\epsilon _2\pi)
\Q(\z '+i\epsilon_3\pi)\Q (\z '+i\epsilon_4\pi)$$
where $\epsilon$ takes values $0,+,-$.
We divide the result into three parts:
\begin{align}
\sum\limits _{l=1}^g \Q(\z)
R_l(Z')\widetilde{S}_l(Z,W)
\Q(\z )\ \Q(\z ')^2=\frac 1{i\pi}\(
\mathcal{A}(\z,\z ')+
\mathcal{B}(\z,\z ')+\mathcal{C}(\z,\z ')\)\non
\end{align}
where
\begin{align}
&\mathcal{A}(\z,\z ')=
\frac {Z '}{Z-Z'}
\bigl\{
\Q(--00)-\Q(++00)\bigr\}\non\\
&\mathcal{B}(\z,\z ')=\frac {Z }{Z-Z'}
\bigl\{ \Q(0++0)-\Q(0--0)+
\Q(0+-0)-\Q(0-+0)
\bigr\}\non
\\ 
&\mathcal{C}(\z,\z ')=
T_{0}\ \bigl\{\Q(-000)-\Q(0+00)
\bigr\}\non
\end{align}

Notice that $\mathcal{A}+\mathcal{B}$ is not singular at $Z=Z'$
as a whole, but since we shall need to consider the items
separately a self-consistent way of understanding
the singularities is to be prescribed.
We shall understand assume that $\zeta $ is
slightly moved to the upper half-plane.
Define the functions:
\begin{align}
&F[b](\z  )=\int\limits _{-\infty }^{\infty }
\frac {Z' }{Z (1+i0)-Z' }
\ b(z' )\ \Q(\z ' )^2d\z ' ,\non\\
&H_{\pm}[b](\z  )=\int\limits _{-\infty }^{\infty }
\frac {Z }{Z (1+i0)-Z' }
\ b(z' )\Q(\z ' \pm  i\pi)\Q(\z ' )d\z ' ,\label{FHH}
\end{align}
These functions satisfy the equations:
\begin{align}
&\delta _{\ga }\bigl(F[b]\bigr)(\z )\hskip .1cm =
\hskip .1cm i\ga\ b(z)\ \Q(\z)^2,\non\\
&\delta _{\ga }\bigl(H_{\pm}[b]\bigr)(\z )=i\ga
\ b(z)\Q(\z \pm  i\pi)\Q(\z ),\label{delta}
\end{align}

Let us return to calculations. Consider, first, the
integral of 
$\mathcal{A}$ rewriting it as follows:
\begin{align}
&
\int\limits _{-\infty}^{\Lambda }d\z  
\int\limits _{-\infty}^{\infty}d\z ' 
\ \mathcal{A}(\z  ,\z ' )=\non\\&=
\int\limits _{-\infty-i\pi}^{\Lambda -i\pi}
\ \Q (\z  )^2F[b](\z  +i\pi)
a(z  )d\z 
-
\int\limits _{-\infty+i\pi}^{\Lambda +i\pi}
\ \Q (\z  ) ^2F[b](\z  -i\pi)
a(z  )d\z 
\non
\end{align}
We want to
move the contours of integration 
by $i\pi $ down in the first integral and by $i\pi $ up
in the second. In order to do that we have to
understand the analytical continuation of $F[b](\z )$. It is easy to
find that for real $\z  $:
\begin{align}
F[b](\z  +i\pi )&=
\int\limits _{-\infty}^{\infty}
\frac {Z' }{Z Q^2-Z' }
b(z' )\ \Q(\z ' )^2d\z ' +\non\\&+i\ga
\sum\limits _{j=1}^{\[\frac \pi \ga \]}b(z q^{-2j})
\ \Q(\z   +i\pi-i\ga j)^2,
\non\\
F[b](\z  -i\pi )&=
\int\limits _{-\infty}^{\infty}
\frac {Z' }{Z Q^{-2}-Z' }
b(z' )\ \Q(\z ' ) ^2d\z ' 
-\non\\&-i\ga
\sum\limits _{j=0}^{\[\frac \pi \ga \]}b(z q^{2j})
\ \Q(\z   -i\pi+i\ga j)^2,\non
\end{align}
Using these formulae we find:
\begin{align}
&
\int\limits _{-\infty}^{\Lambda }d\z  
\int\limits _{-\infty}^{\infty}d\z ' 
\ \mathcal{A}(\z  ,\z ' )=
\label{A=}\\ &=
 -\int\limits _{\Lambda}^{\Lambda +i\pi}
\bigl(
\ \Q (\z  )^2F[b](\z  -i\pi)  +
\ \Q (\z  -i\pi) ^2F[b](\z )\bigr)
a(z  )d\z -\non\\&-
\int\limits _{-\infty}^{\infty}\int\limits _{-\infty}^{\infty}
\frac{Z Z' (Q^2-Q^{-2})\ a(z )b(z' )}
{(Z Q^2-Z' )(Z Q^{-2}-Z' )}
(\Q(\z  )\Q(\z ' ))^2d\z  d\z ' 
-\non\\&-i\ga
\int\limits _{-\infty}^{\Lambda }
\ a(z )b(z )
(\Q(\z   )\Q(\z  -i\pi ))^2d\z  
-\non\\
&-i\ga 
\sum\limits _{j=1}^{\[\frac \pi \ga \]}
\int\limits _{-\infty}^{\infty }
\bigl(
a(z )b(z q^{-2j})+b(z )a(z q^{-2j})\bigr)
(\Q(\z   )\Q(\z   +i\pi-i\ga j))^2
d\z  
+
o(1)\non
\end{align}
where $o(1)$ comes from replacing the upper limits
of rapidly converging integrals 
in the last line from $\Lambda$ to $\infty$.
We also moved contours of integration in some
of these integrals.
The formula (\ref{A=}) is written in such a way that
the symmetry with respect to replacing $a\leftrightarrow b$
is obvious everywhere except the first integral
in the RHS. Our goal is to calculate (\ref{gh}) where
the antisymmetrization is performed with 
respect to $a$ and $b$. So, only the first integral
in the RHS is relevant.

Now let us consider the integral of $\mathcal{B}$ which can be rewritten 
as follows:
\begin{align}
\int\limits _{-\infty}^{\Lambda}d\z  
\int\limits _{-\infty}^{\infty}d&\z ' 
\ \mathcal{B}(\z  ,\z ' )=\non\\=&
\int\limits _{-\infty+i\pi}^{\Lambda +i\pi}a(z  )
\Q(\z  -i\pi)\Q(\z )H_+[b](\z -i\pi)
-\non\\ 
-&\int\limits _{-\infty-i\pi}^{\Lambda -i\pi}a(z  )
\Q(\z )\Q(\z +i\pi)H_-[b](\z +i\pi)d\z
+\non\\+&
\int\limits _{-\infty}^{\Lambda }a(z  )
\Q(\z )\bigl(\Q(\z +i\pi)H_-[b](\z )-
\Q(\z  -i\pi)H_+[b](\z )\bigr)d\z \non
\end{align}
In the first two integrals in RHS we would like
to move the contour of integration to the real axis.
The result is
\begin{align}
&\int\limits _{-\infty}^{\Lambda}d\z  
\int\limits _{-\infty}^{\infty}d\z ' 
\ \mathcal{B}(\z  ,\z ' )=\int\limits _{-\infty}^{\Lambda }
\mathcal{X}[b](\z  )a(z  )d\z +\non\\&
+\int\limits _{\Lambda}^{\Lambda +i\pi}a(z  )
\Q(\z )
\Q(\z -i\pi)\bigl(H_-[b](\z )+
H_+[b](\z -i\pi )\bigr)d\z \non
\end{align}
where
\begin{align}
\mathcal{X}[b](\z  )=&
\Q(\z  -i\pi)H_+[b](\z -i\pi)
-\Q(\z )\bigl(\Q(\z +i\pi)H_-[b](\z +i\pi)
+\non\\
+
&\Q(\z +i\pi)H_-[b](\z )
 -\Q(\z  -i\pi)H_+[b](\z )
\bigr)\label{intX}
\end{align}

This is the place to discuss asymptotic of $H_{\pm}[b]$
and $F[b]$. 
Due to (\ref{delta})  these asymptotic must be related to
$\delta _{\ga }^{-1}(b(z)\varphi _{\epsilon}(\z)
\varphi _{\epsilon '}(\z ))$ as they were defined in the 
Section 2. 
There is, however, a subtlety:  the
asymptotic may differ by
quasiconstants: $i\ga $-periodical functions.
For $H_{\pm}[b]$ the situation is simple,
one finds that for
$0\le \text{Im}(\z )\le \pi $
\begin{align}
&H_{-}[b](\z )\simeq H_{+}[b](\z-i\pi )=\label{asH}\\&=i\ga
\(\delta _{\ga }^{-1}\bigl(
b(z)\varphi _-(\z )\varphi _+(\z )\bigr)
-b(z)\varphi _-(\z )^2\)
\(1+O(e^{-2(g+1)})\)\non
\end{align}
The asymptotics of $F[b](\z )$ in the strip
$-\pi\le \text{Im}(\z )\le \pi $ can be found:
\begin{align}
F[b](\z)&=\sum\limits _{k=1}^g
Z^{-k}\langle b,R_k\rangle+\label{asF}\\&+i\ga
\ Z^{-(g+1)}\(2\delta _{\ga }^{-1}
(b(z)\varphi _-(\z )\varphi _+(\z ))-
b(z)\varphi _-(\z )^2\)
\(1+O(e^{-2(g+1)})\)
\non
\end{align}
 
Using the asymptotical formulae (\ref{asH}) we find that
$$\mathcal{X}[b](\z )=-i\ga\ b(z)\(\Q(\z)\Q(\z -i\pi )\)^2+O(z^{-1})$$
Recall that we have finally to anti-symmetrize. The above
formula guaranties rapid convergence of the integral:
$$\int\limits _{-\infty}^{\Lambda}
\(a(z)\mathcal{X}[b](\z )-b(z)\mathcal{X}[a](\z )\)d\z
$$
So, the limit $\Lambda \to\infty$ can be taken for this
integral separately. It can be shown further that 
$$\int\limits _{-\infty}^{\infty}
\(a(z)\mathcal{X}[b](\z )-b(z)\mathcal{X}[a](\z )\)d\z=0
$$

Let us summarize taking into account the integrals
containing $P_l$ in (\ref{pred}):
\begin{align}
&\sum\limits _{l=1}^g \bigl(
\langle a, S_l\rangle
\langle b, R_l\rangle 
-\langle  a, R_l        \rangle
\langle  b, S_l         \rangle\bigr)=
\lim _{\Lambda\to\infty}\frac 1 {\pi i}
\int\limits _{\Lambda}^{\Lambda +i\pi}a(z)\times\non\\
&\times\bigl\{
\Q(\z)\Q(\z -i\pi)\bigl(
H_-[b](\z)+H_+[b](\z-i\pi)-\sum\limits _{k=0}^gT_k
\langle b, R_k\rangle\bigr)-
\non\\&-
\Q(\z-i\pi )^2\bigl(
F[b](\z)-\sum\limits _{k=1}^g
Z^{-k}\langle b, R_k\rangle
\bigr)-\non\\&-
\Q(\z )^2\bigl(
F[b](\z-i\pi)-\sum\limits _{k=1}^g
Z^{-k}Q^{2k}\langle b, R_k\rangle
\bigr)\bigr\}d\z -(a\leftrightarrow b)\label{finint}
\end{align}

Using the asymptotics (\ref{asH},\ref{asF}) and the fact that
$\Res (a)=\Res (b)=0$ one finds
\begin{align}
&\sum\limits _{l=1}^g\bigl(
\langle a, S_l\rangle
\langle b, R_l\rangle 
-\langle  a, R_l        \rangle
\langle  b, S_l         \rangle\bigr)= 
\lim _{\Lambda\to\infty}
\ \frac {\ga }{\pi}\int\limits _{\Lambda}^{\Lambda +i\pi}
a(z)\times\label{konec}\\ \times&\bigl(
2\varphi  (\z) _+\varphi _-(\z)
\delta _{\ga}^{-1}(b(z)\varphi _+(\z)\varphi _-(\z))+
b(z)\bigl(\varphi _+(\z)
\varphi _-(\z)\bigr)^2\bigr)d\z =a\circ b \non
\end{align}
where the formula (\ref{pair1}) for pairing has been used.

\hfill{\bf QED}\vskip .2cm

The Riemann bilinear identity can be reformulated in the following fashion.
\newline
{\bf Corollary 1.}
Consider the $2g\times 2g$-matrix 
$$ \mathcal{P}_{2g}=\begin{pmatrix}
\langle r,R\rangle ,&
\langle r,S\rangle\\
\langle s, R\rangle ,&
\langle s,S\rangle 
\end{pmatrix}
$$
then 
\begin{align}
&\mathcal{P}_{2g}\in Sp(2g)\label{sp2g}
\end{align}

Until now we have considered only first and second
king differentials, but actually the third kind differential
$r_{g+1}$ can also be taken into account.
In the proof of the Theorem we did not use the requirement
$\Res (a)=0$ up to the very last transformation from (\ref{finint})
to (\ref{konec}). Suppose $a=r_{g+1}$ and $b=r_k$,
$k=1,\cdots ,g$. In the formula (\ref{finint}) we have the expression
\begin{align}
H_-[b](\z)+H_+[b](\z-i\pi)-\sum\limits _{k=0}^gT_k
\langle b, R_k\rangle\simeq 2i\ga\ \delta _{\ga}^{-1}
\(b(z)\varphi _+(\z)\varphi _-(\z)\)\label{HH=d}
\end{align}
Usually $\delta _{\ga}^{-1}$ is defined up to a constant
which was until now irrelevant. However,  in the formula
(\ref{HH=d}) this constant can be calculated:
$$C=\langle b, R_{g+1}\rangle$$
where the requirement $b=r_k$ was important for 
convergence of integrals. 
Obviously this is the constant $C$ which is important
for calculation of the integral containing (\ref{HH=d}) because
$a=r_{g+1}$ has residue. The remaining integrals as well as
the term $(a\leftrightarrow b)$ do not contribute. So, we find
for $b=r_k$:
\begin{align}
\sum\limits _{l=1}^g\bigl(
\langle r_{g+1}, S_l\rangle
\langle b, R_l\rangle 
-\langle  r_{g+1}, R_l        \rangle
\langle  b, S_l         \rangle\bigr)=
\langle b, R_{g+1}\rangle
\label{rg+1b}
\end{align}
With somewhat more elaborate
calculations we find the same result for $b=s_k$ for which
$\langle b, R_{g+1}\rangle$ is given by usual formula (\ref{sR}).

\section{Classical limit.}

In this section we show that
in the classical limit the deformed Abelian differentials
and integrals
reproduce their classical counterparts. But we have to
start with some more explicit formulae concerning the
classical limit of solutions to Baxter equation.

Consider the hyper-elliptic algebraic curve $X$:
$$w^2-t(z)w+1=0$$
where the real polynomial $t(z)$ with real positive zeros:
$$t(z)=\prod\limits _{k=1}^{g+1}(z-x_k)$$
We assume that $t(0)>2$ and 
 zeros of the
polynomial $t(z)^2-4$ are real positive. 
We shall denote by $y_j$ the branch points:
$$t(z)^2-4=\prod\limits _{j=1}^{2g+2}(z-y_j)$$
Obviously in the above conditions we have
$$y_{2k-1}<x_k<y_{2k}$$
Introduce the notation
$$\xi _j=\half \log{y_j}$$
The real axis appears to be divided into intervals:
$$
I_k=\[\xi _{2k},\xi _{2k+1}\] ,\quad k=0, \cdots g+1 ,
$$
where $\xi _0=-\infty$,  $\ \xi _{2g+3}=\infty$.
$$ 
J_k=\[\xi _{2k-1},\xi _{2k}\],\quad k=1, \cdots g+1 
$$
Let us realize the Riemann surface as follows.
In the plane of $\z$ consider the strip $-\pi<\text{Im}(\z )< \pi$
with cuts along the
intervals $I_k$.
We identify the following segments:
\begin{align}
&I_k+i0=I_k+i\pi -i0,\quad I_k-i0=I_k-i\pi +i0,\non\\&
J_k+i\pi-i0=J_k-i\pi +i0,\non
\end{align}
The segments $[\infty ,\infty +i\pi]$, $[\infty-i\pi ,\infty ]$  are
contructed
to points $\infty ^{+}$, $\infty ^{-}$, the segments 
$[-\infty ,\infty+i\pi]$,  $[-\infty -i\pi ,\infty ]$ 
are contructed to the points  $0^+$ and $0^-$
(they correspond to two points over $z=0$
if the surface is realized as covering of $z$-plane),
the latter two points
do not play any special role in our considerations.
The a-cycle $\alpha _k$
goes around  the cuts $I_k$ ($k=1,\cdots ,g$), the b-cycle
$\beta _k$ goes in the upper half-strip form a point on 
$I_k$ to identified point on $I_k+i\pi$.
\vskip 1cm
\hskip -0.5cm  
\epsffile{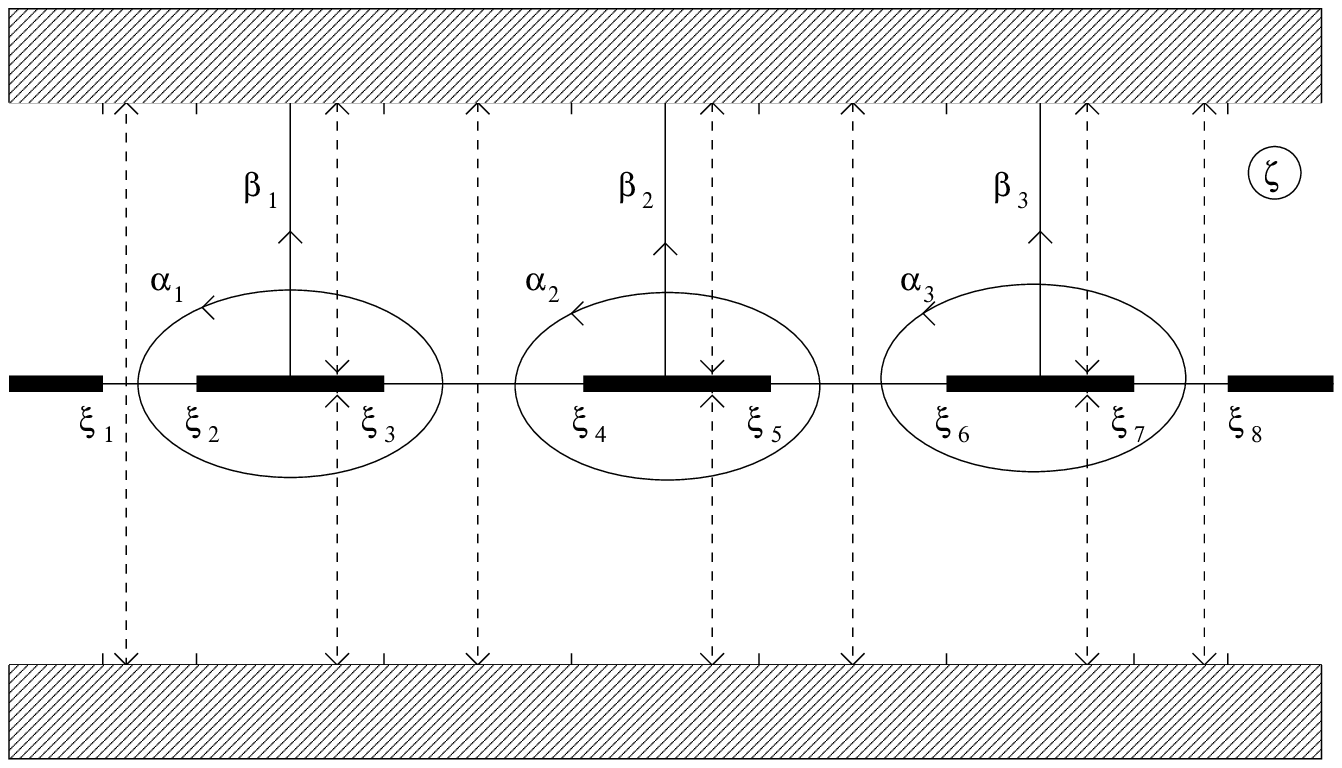}
\vskip 0.2cm
\noindent

Consider $\log w$. It is not a single-valued function on
the surface. Actually $d\log w$ is quite special Abelian
differential:
it is of the third kind with simple poles at $\infty ^{\pm}$
and with vanishing a-periods. So, $\log w$ is single-valued
on the plane with cuts, its imaginary part is constant
along the cuts:
\begin{align}
&\text{Im}(\log w)=-\pi k \quad\text{at} \quad I_k\pm i0,\non\\
&\text{Im}(\log w)=\hskip .35cm\pi k\quad\text{at} \quad
I_k\pm  (i\pi-i0),\non
\end{align}
Away from the cuts the quasi-classical solution to the
Baxter equation is given by
\begin{align}
\Q^{cl} _+(\z)=\(t(z)^2-4\)^{-\frac1 4}\exp
\left\{\frac 1 {i\ga}\int\limits _{-\xi _1}^\z \log w\ d\z +
\frac {\pi i} 4\right\}
\label{qcQ}
\end{align}
where $\(t(z)^2-4\)^{\frac1 4}$ is defined to be positive on
$I_k+i0$.  The function (\ref{qcQ}) is 

Actually we have two solutions: one of them is (\ref{qcQ})
and another with minus in the exponent. However, this
second solution is small everywhere except the
intervals $I_k$  where oscillations take place.
Inside these intervals we have to take for $\Q^{cl}(\z)$ the
sum of values of (\ref{qcQ}) on the upper and
lower banks:
\begin{align}
\Q^{cl}(\z)=\Q^{cl}_+(\z+i0)+\Q^{cl}_+(\z-i0),\quad 
\z
\in I_k
\end{align}
The definition of $\Q^{cl}(\z)$ in $I_0$ is obvious,
for the rest of $I_k$ a
simple calculation gives:
\begin{align}
&\text{for}\quad \z\in I_k,\ k=1,\cdots ,g+1:\non\\
&\Q^{cl}(\z )= C_k \ e^{-\frac \pi \ga j\z}
\ \left|t(z)^2-4\right|    ^{-\frac 1 4}
2\cos \left\{\frac 1{\ga }\int\limits _{\xi _{2k}}^\z \log|w|
d\z +\frac {\pi i} 4\right\}\label{QinI}
\end{align}
where
\begin{align}
&C_k=\exp \left\{ \frac {\pi}{\ga}\sum\limits _{l=1}
^{k}\sigma _l     \right\}\non
\end{align}
where
\begin{align}
\sigma _l=\xi _{2l-1}+\frac 1{\pi i} \int\limits
_{\xi_{2l-1}}^{\xi _{2l}}
\(\log w+\pi i (l+1)\)d\z
\end{align}
By simple contour integration one finds:
\begin{align}
&\sum\limits _{l=1}^{g+1}\sigma_l=0\label{X=1}
\end{align}
Inside the interval $J_{l}$ 
the function
$\log w$ is pure imaginary. Moreover, the real in this
interval function 
$$\frac 1{\pi i} 
\(\log w+\pi i l\)$$
decreases monotonically from $1$ to $0$.
So, we have an important inequality:
\begin{align}
\xi _{2l-1}<\sigma _l<
\xi _{2l}\label{ysy}
\end{align}

Let us understand the meaning of $\sigma _l$.
Recall that in the quantum case duality holds which implies that
\begin{align}
\Q(\z +i\pi)+\Q(\z -i\pi)=T(Z)\Q(\z )\label{QQTQ}
\end{align}
where $T(Z)$ is a polynomial
$$T(Z)=\prod\limits _{l=1}^{g+1}(Z-X_l)$$
Let us show that quasi-classically
\begin{align}
&X_l
\simeq 
e^{\frac {2\pi}{\ga}\sigma _l}
\label{X}
\end{align}

Indeed,  take $\zeta\in I_k$ and consider
$\Q^{cl}(\z +i\pi)$ and $\Q^{cl}(\z -i\pi)$.
They are understood as analytical continuations
respectively of $\Q^{cl} _+(\z +i0)$ and $\Q^{cl} _+(\z -i0)$.
After some calculation we find:
\begin{align}
&\Q^{cl}(\z +i\pi)=(-1)^{g+1-k }Z^k
C_k^{-2}\ \Q^{cl}_+(\z+i0),
\non\\
&\Q^{cl}(\z -i\pi)= (-1)^{g+1-k}Z^k
C_k^{-2}\ \Q^{cl}_+(\z-i0),\quad \z\in I_k
\label{+pii}
\end{align}
the only contribution which needs explanations is $(-1)^{g+1-k }$
which comes from branch points of $(t^2-t)^{\frac 1 4}$.
Quasi-classically inside $I_k$ one has
$$ X_i\ll Z\ll X _j,\quad i \le k< j$$
So, 
putting together (\ref{+pii}) and (\ref{QinI}) we prove that
the equation (\ref{QQTQ}) holds in $I_k$ quasi-classically.


Let us consider the classical limit of 
deformed Abelian differentials.
Recall that the deformed Abelian differentials 
are defined by polynomials $a\in \text{v}$. For corresponding
Abelian differential we want to take
$$\omega =\frac {a(z)}{z\sqrt{t(z)^2-4}}\ dz$$
but one should take limit $\ga\to 0$ of $a(z)$ if the
latter depends explicitly upon $\ga$.

The polynomials $r_k(z)=z^{k}$ 
do not depend on $\ga$ while for
$s_k$ one finds:
$$\lim _{\ga\to 0} s _k(z)=
t(z)\ z\frac d {dz}\[z^{-k}t(z)\]_>$$
The differentials
$$\omega _k =\frac {z^{k}}{z\sqrt{t(z)^2-4}}\ dz$$
for $k=1,\cdots ,g$ are of the first kind.
They vanish at $\infty +i0$ as follows:
\begin{align}
\omega _k=
\frac {z^{k-1}}{t(z)}\(1+O(z^{-2(g+1)})\)dz
\label{inf1}
\end{align}

The differential 
$$\omega _{g+1}=\ \frac {z^{g}}{\sqrt{t(z)^2-4}}\ dz$$
is of the third king, it behaves at $\infty +i0$ as follows
$$\omega _{g+1}=\frac{dz}z\(1+O(z^{-1})\) $$
The differentials 
$$\widetilde{\omega}_k=t(z)\ \frac d {dz}\[z^{-k}t(z)\]_>
\frac {1}{\sqrt{t(z)^2-4}}\ dz$$
are of the second kind with the following singularities at
$\infty ^{\pm}$:
\begin{align}
\widetilde{\omega}_k=
d\[z^{-k}t(z)\]_>\(1+O(z^{-2(g+1)})\)\label{inf2}
\end{align}

The differentials $\omega _k$, $\widetilde{\omega}_k$
do not have residue on the surface. For such differentials
we define pairing:
$$\omega \circ \omega '=\sum \text{res}
\ \bigl(\omega d^{-1}\omega '\bigr)$$
The 
only singularities of  $\omega _k$, $\widetilde{\omega}_k$
are situated at the points $\infty ^{\pm}$. From
(\ref{inf1}) and (\ref{inf2}) it is clear that 
$\omega _k$, $\widetilde{\omega}_k$ constitute
a canonical basis:
$$\omega _k\circ\omega _l=
\widetilde{\omega}_k   \circ\widetilde{\omega}_l=0,\quad
\omega _k\circ\widetilde{\omega}_l=\delta _{k,l}
$$
All these formulae actually explain the terminology used 
before. Now we have to do some analysis in order to
calculate classical limit of deformed Abelian integrals.

Consider, first, the pairing
$$\langle a, R_k\rangle
=\int\limits _{-\infty}^{\infty}a(z)\ Z^{k}\ \Q(\z)^2d\z$$
for $k=1,\cdots g$. For
small $\ga$ the integral  rapidly converges.
What we need to know is the quasi-classical
behaviour of $\Q(\z )^2$. 
We have seen already that in the intervals $I_k$ the function
$\Q^{cl} (\z)$ rapidly
oscillates.  In the intervals $J_k$ the function
$\Q^{cl}(\z)$ is real positive.  Up to rapidly oscillating part
which is denoted by $\cdots$ we can
present $\Q^{cl}(\z )^2$ for real $\z $ in universal way:
\begin{align}
&\Q^{cl}(\z)^2=\frac 1 {\sqrt{t(z)^2-4}}
\exp \left\{\frac 2 \ga \int\limits _{-\infty}^{\z}\arg w\ d\z\right\}+\cdots
\end{align}
The function $\arg w$  decreases monotonously from 
$0$ to $-\infty$ when $\z $ goes from $-\infty$ to $\infty$.
It is constant  equal to $-\pi l$ in the interval $I_l$.
Under the integral we have the function
$\Q^{cl}(\z )^2Z^{k}$.
The two multipliers compete:  $\Q^{cl}(\z )$ decreases
and  $Z^{k}$ grows. The result is that in quasi-classical
limit
the integrand is everywhere exponentially small with
$\frac 1 \ga$ in exponent  with respect to 
its values in $I_{k}$.
Now using explicit formula 
(\ref{QinI}) we find:
\begin{align}
&\langle a, R_k\rangle
\simeq 
C_k^2
\int\limits _{\alpha _{k}}
\frac{a(z)}{z\sqrt{t(z)^2-4}}\ dz\label{aper}
\end{align}

Now we turn to the pairings $\langle a, S _k\rangle $. 
They look much more complicated, but the final calculation
will be even simpler. In order to simplify the situation let
us use the following trick. Up to now we always considered
$\ga $ in generic position, when $\ga $ is a rational multiple of
$\pi $ everything becomes much more complicated,
literally some of our formulae loose meaning. It is not 
so bad, but some additional work is needed.
However, the limit $\ga \to 0$ can be taken along
some special direction, for example, we can put
$$\ga =\frac {\pi}{n}, \quad n\in \mathbb{N}$$
and take the limit $n\to \infty$. For sufficiently large $n$ 
the problems mentioned above disappear in every
particular formula. We insist on considering this values 
of $\ga $ because for them the formulae for pairings
simplify a lot.
There are two important simplifications for
these values of $\ga$: first, $Q^2=1$, second,
$Z$ becomes $\pi$-periodic. Recall the regularizations
(\ref{rS}) and (\ref{sS2}). For $n\gg 1$ (small $\ga $)
we can use the formula (\ref{simpl}) for
$\langle s_l, S_k\rangle$.
In other words 
we can take the same
formula for $\langle a, S_k\rangle $  if $a=r_l$ or $a=s _l$
\begin{align}
&\langle a, S_k\rangle =\int\limits _{-\infty} ^{\Lambda}\Q(\z)a(z)
\widetilde{S}_k(Z,W)\Q(\z)d\z+\non\\&
+\int\limits _{\Lambda} ^{\Lambda +i\pi}\Q(\z)a(z)
P_k(Z,W^*\hskip -.1cm,W)\Q(\z)d\z
-\int\limits _ {\Lambda}^{\infty}\Q(\z)a(z)
\widetilde{S}_k^{\ -}    ( Z,W)        \Q(\z)d\z\non
\end{align}
Notice that for the values of $\ga$ under consideration
\begin{align}
&\Q(\z)\widetilde{S}_k(Z,W)\Q(\z)=
\ \delta _{\pi}\([Z^{-k}T(Z)]_>\Q(\z-i\pi)\Q(\z)\)\non\\
&\Q(\z)\widetilde{S}_k^{\ -}    ( Z,W) \Q(\z) =
\delta _{\pi}\([Z^{-k}T(Z)]_<\Q(\z-i\pi)\Q(\z)\)
\non
\end{align}
So, we find the following beautiful result:
\begin{align}
&\langle a, S_k\rangle =
\int\limits _{\Lambda} ^{\Lambda +i\pi}a(z)Z^{-k}
\bigl(\Q(\z)\Q(\z)-\Q(\z-i\pi)\Q(\z+i\pi)\bigr)
d\z
\label{aSn}
\end{align}
This is exact result for $\ga =\frac \pi n\ $! It is a nice exercise
to show that for these values of $\ga$ the RHS of
(\ref{aSn}) does not depend on $\Lambda$

Now we are ready to take the classical limit. First, notice that quasi-classically
$$|\Q^{cl}(\z)|^2\ll |\Q^{cl}(\z-i\pi)\Q^{cl}(\z+i\pi)|$$ 
for $0\le\text{Im}(\z)\le\pi$, so, we can neglect the first
part of the integrand. Now let us use the freedom in
definition of $\Lambda$ putting it
inside the interval $I_{k}$. Use the formulae
(\ref{+pii}) in order to estimate 
$\Q^{cl}(\z-i\pi)\Q^{cl}(\z+i\pi)$ for $\z\in I_{k}$:
\begin{align}
\Q^{cl}(\z-i\pi)&\ \Q^{cl}(\z+i\pi)=\non\\=
&Z^{2k}\ C_k^{-4}
\ \Q^{cl} _+(\z-i0)\ \Q^{cl} _+(\z+i0)=
Z^{k}\ C_k^{-2}
\label{asQQ}
\end{align}
The leading term of asymptotics (\ref{asQQ}) continues
analytically to the strip $0\le\text{Im}\z\le \pi$. Since
the interval $[\Lambda, \Lambda +i\pi]$ for $\Lambda \in
I_{k}$ is the b-cycle $\beta _k$ , thus
\begin{align}
&\langle a, S_k\rangle \simeq
C_k^{-2}
\int\limits _{\beta _{k}} \frac {a(z)}
{z\sqrt{t(z)^2-4}}
\ dz
\label{bper}
\end{align}

Now it is clear that the Riemann bilinear relation for
deformed Abelian differential turn in the classical limit
into usual Riemann bilinear relation: for two differentials
$\omega$, $\omega '$ which can be singular
only at $\infty^{\pm}$ and do not
have residues the relation holds:
\begin{align}
\sum\limits _{l=1}^g\(
\ \int\limits _{\alpha _l}\omega
\ \int\limits _{\beta _l}\omega '-
\int\limits _{\alpha _l}\omega '
\ \int\limits _{\beta _l}\omega \)=
\omega\circ \omega '
\end{align}

On the affine curve 
$$X_{\text{aff}}=X-\infty ^{\pm}$$
one can consider two additional cycles: 
the first one ($\sigma $) goes around
$\infty ^+$, the second one ($\rho$) is non-compact: 
it goes from $\infty ^-$ t0 $\infty ^+$. Obviously
$$\sigma\circ\rho =1$$
The integrals over $\sigma$ are well defined:
\begin{align}
&\int\limits _{\sigma}\omega =\text{res}_{\infty ^+}(\omega)
\label{so}
\end{align}
The second cycle is not included into usual
homology theory, but we can define integrals over
$\rho$ of all first and second kind differentials.
For the first kind differential the integral is well defined
from the very beginning. For the second kind one we shall
do the following:
\begin{align}
\int\limits _{\rho}\widetilde{\omega}_k\equiv
-\int\limits _{\rho} t(z)\(\frac d{dz}\[z^{-k}t(z)\]_<+4kz^{-k-1}\)
\frac {dz}{\sqrt{t(z)^2-4}}
\label{ro}
\end{align}
which means that we have subtracted from 
$\widetilde{\omega}_k$ the exact form 
$d\bigl(z^{-k}\sqrt{t(z)^2-4}\bigr)$.
We consider $\sigma$ and $\rho$ as classical limit
of respectively $S_{g+1}$ and $R_{g+1}$. The formulae
(\ref{so}), (\ref{ro}) reproduce classical
limit of corresponding quantum formulae.
Let us consider the classical limit of the
relation (\ref{rg+1b}):
\begin{align}
\sum\limits _{l=1}^g\(
\ \int\limits _{\alpha _l}\omega _{g+1}
\ \int\limits _{\beta _l}\omega -
\int\limits _{\alpha _l}\omega
\ \int\limits _{\beta _l}\omega _{g+1} \)=
\int\limits _{\rho}\omega
\end{align}
If $\omega $ is of the first kind this is obvious form
of Riemann bilinear relation with third kind differential
$\omega _{g+1}$: in the RHS we have difference
of values of primitive function of $\omega $ at the points
where $\omega _{g+1}$ has simple poles.
For second kind differentials we subtract the exact
form as it is explained above, this does not change the 
periods of $\omega$ in the LHS. In the RHS we have
the integral over $\rho$ defined by (\ref{ro}) plus
contribution from residue at $z=0$ where 
$d\bigl(z^{-k}\sqrt{t(z)^2-4}\bigr)$ is singular.
However $r_{g+1}$ vanishes as $z^{g+1}$ at
$z=0$ so the second contribution disappears.

\end{document}